\documentclass[aps,prl,twocolumn, groupedaddress]{revtex4}
\usepackage{amssymb}
\usepackage{graphicx,amsmath}

\begin{document}

\title{Nonequilibrium quasiparticles and $2e$ periodicity in
single-Cooper-pair transistors\footnote{%
Contribution of NIST; not subject to copyright in the U.S.}}
\author{J. Aumentado}
\email[]{jose.aumentado@boulder.nist.gov}
\author{Mark W. Keller}
\author{John M. Martinis}
\affiliation{National Institute of Standards and Technology, Boulder, CO 80305-3337}
\author{M.H. Devoret}
\affiliation{Department of Applied Physics, Yale University, New Haven, CT 06520-8284}
\date{\today}

\begin{abstract}
We have fabricated single-Cooper-pair transistors in which the spatial 
profile of the superconducting gap energy was controlled by oxygen doping. The profile dramatically affects the switching current \textit{vs.} gate voltage curve
of the transistor, changing its period from $1e$ to $2e$. A model based on nonequilibrium quasiparticles in the leads explains our results, including the surprising observation that even devices with a clean $2e$ period are ``poisoned'' by small numbers of these quasiparticles.
\end{abstract}

\maketitle

%\preprint{}




%\thanks{}



%\keywords{}


Coherent superpositions of charge states can be prepared and manipulated in
circuits made of ultrasmall superconducting junctions \cite{nakamuratsai,vionscience}. This phenomenon may enable the construction of 
solid-state qubits based on charge states, as well as a quantum current standard whose speed is not limited by the stochastic nature of incoherent tunneling. In both cases, tunneling of unpaired quasiparticles (QPs) causes decoherence and may limit operation to impractically short timescales. Reducing this ``QP poisoning'' requires a detailed understanding at a fundamental level.

\begin{figure}[tbp]
\includegraphics[scale=0.5]{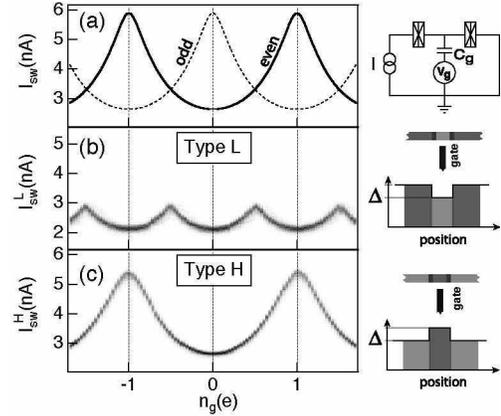} %0.3
\caption{(a) Schematic representation of $I_{sw}^{even/odd}$. Single-Cooper-pair transistor circuit shown at right. (b)--(c)
Grayscale plots of $I_{sw}(n_{g})$ for co-deposited type~L and type~H SCPTs
at $T=30$~mK, measured with $\dot{I}=1$~$\protect\mu $A/s. The corresponding gap profiles are shown on the right.}
\label{fig1}
\end{figure}

The single-Cooper-pair transistor (SCPT), shown in Figure~\ref{fig1}(a), consists of a micrometer-sized island that has a capacitive gate electrode and is probed by two Josephson junctions with areas $\sim$(100~nm$)^2$. The island charging energy is modulated by the gate according to $E_{C}^{n}(n_{g})=E_{C0}(n-n_{g})^{2}$, where $E_{C0}\equiv e^{2}/2C_{\Sigma }$, $e$ is the electron charge, $C_{\Sigma }$ is the total island capacitance, $n_{g}\equiv C_{g}V_{g}/e$ is the normalized gate polarization, and $n$ is the integer number of excess charges on the island. This Coulomb energy suppresses fluctuations in the island charge and causes the Josephson energy $E_{J}$ to vary with $n_{g}$ through charge-phase duality \cite{joyezprl}. This effect is strongest when $E_{C0}\gtrsim E_{J}\gg kT$, where $k$ is the Boltzmann constant and $T$ is the temperature. In a current-biased configuration, the modulation of $E_{J}$ manifests itself in the current $I_{sw}$ at which the SCPT switches from the ``supercurrent branch'' near zero voltage to the ``voltage state'' \cite{kautzprb}. $I_{sw}(n_{g})$ is maximized when charge states differing by one Cooper pair are degenerate, yielding a $2e$ periodic modulation in $n_{g}$. Without QP tunneling, each $2e$ interval of $I_{sw}(n_{g})$ has a single peak, decreasing monotonically to a valley on either side.  This shape, which we call ``clean'' $2e$ modulation, is illustrated in Fig. \ref{fig1}(a) for parity labeled ``even'' or ``odd'' depending on $n$. Intermittent tunneling of QPs during a measurement causes random changes in parity and results in a more complicated modulation containing secondary peaks and valleys. Thus clean $2e$ modulation has often been viewed as an indication that an SCPT is free of QP poisoning.

%Fig 1 stays the same

Early studies of SCPTs (many unpublished) showed a $1e$ period, indicating strong QP poisoning, even when thermal QPs were suppressed. Nonequilibrium QPs in the leads were thought to be responsible, leading one group to put normal metal leads close to their junctions to filter QPs \cite{joyezprl}. This yielded a clean $2e$ period in almost all devices and appeared to resolve the issue. However, other experiments revealed a variety of unexplained results, including a $2e$ period without QP filters \cite{tuominenprl,wellstoodprl}, $2e$ in some devices and $1e$ in others even though all devices shared the same design, fabrication process, and measurement setup \cite{eilesprb,eilesthesis}, and finally poisoning that \emph{increased} as $T$ was lowered in some samples \cite{vanderwalthesis}. In this sense, QP poisoning in SCPTs has remained a puzzle, which motivated us to create devices that were controllably either $1e$ or $2e$ and understand in detail how QPs behave in both types.

SCPT fabrication typically involves a double-angle Al deposition, with a junction oxidation step between layers, that creates the leads in one deposition and the island in the other. Since the process is completed without breaking vacuum, the Al films are usually assumed to be identical, but minor variations in vacuum conditions may cause differences in film disorder that can affect the superconducting gap energy $\Delta$. Thus an SCPT may have a lead gap $\Delta _{\ell}$ that is different from the island gap $\Delta _{i}$. We have exaggerated this effect by oxygen-doping the first deposition \cite{chiclarkeprb}, allowing us to co-fabricate SCPTs in which $\delta\Delta\equiv\Delta _{l}-\Delta _{i}$ is both positive (type L) and negative (type H) (see Fig.~\ref{fig1}(b) and (c)). Starting from a base pressure $\simeq$3$\times$10$^{-7}$~mbar, we deposited the first Al layer (20~nm thick, 0.4~nm/s) while flowing O$_{2}$ gas to raise the pressure to $\simeq$5$\times $10$^{-6}$~mbar. We then oxidized at room temperature in 130~mbar of O$_{2}$ for 5~minutes. After rotating the substrate to the second angle and pumping back to base pressure, we deposited the second layer (30~nm thick, 0.4~nm/s) with no O$_{2}$ added. Measurements of individual films showed the oxygen-doped deposition always had a higher critical temperature $T_{c}$ and thus a larger $\Delta$. We note that our devices have \emph{no QP filters near the junctions}, however the Al leads connect to Au/Ti pads $\simeq10~\mu$m away.

%We approach the issue of nonequilibrium QP poisoning by altering the spatial profile of the superconducting gap energy $\Delta$, exaggerating an effect that may occur unintentionally in typical SCPTs. The most common method of SCPT fabrication utilizes a double-angle Al evaporation, creating the leads in one deposition and the island in another, with a junction oxidation step between layers \cite{fultondolan}. Because both depositions are performed without breaking vacuum, it is usually assumed that the Al films are identical. However, minor variations in vacuum conditions may lead to differences in the film disorder which can affect $\Delta$. Thus an SCPT produced in this manner may have a lead gap $\Delta _{\ell}$ that is different from the island gap $\Delta _{i}$. 

%We co-fabricated SCPTs in which $\delta\Delta\equiv\Delta _{l}-\Delta _{i}$ was intentionally made both positive (type L) and negative (type H) using mirror symmetry in the e-beam lithography patterns (see Fig.~\ref{fig1}(b) and (c)). To increase $|\delta\Delta|$ in a controlled manner, we oxygen-doped the first deposition \cite{chiclarkeprb}. 

All measurements were performed in a dilution refrigerator with a base temperature of 25~mK using coaxial leads with microwave filtering at both 4~K and the mixing chamber. The devices were mounted inside an rf-tight copper box and measured using a two-probe, current-biased configuration. The bias current was ramped using a linear voltage ramp applied across a 10~M$\Omega$ resistor (at 4~K) in series with the SCPT. In this configuration the current-voltage characteristic was hysteretic with a distinct supercurrent branch at zero bias. We measured $I_{sw}$, the current at which the SCPT switched to the voltage state, by cycling through the hysteresis loop $10^{3}$--$10^{4}$ times at each value of gate voltage. After each switch, we returned to the supercurrent branch and allowed the system to equilibrate for at least 1~ms before starting the next ramp (longer times did not change our results). In this manner, we acquired a histogram of $I_{sw}$ at each value of $n_{g}$, allowing us to study the \emph{distribution} of switching currents rather than only the mean as in most previous studies of SCPTs.

Here we focus on two co-deposited devices which are representative of the L and H gap profiles shown in Fig.~\ref{fig1} and which are otherwise nearly identical. We determined the gap energies of the first and second depositions from the measured critical temperatures, yielding $\Delta_{1}=246~\mu$eV ($T_{c1}=1.63$~K) and $\Delta_{2}=205~\mu$eV ($T_{c2}=1.36$~K). We determined the charging energies using the asymptotic current-voltage characteristics \cite{wahlgrenprb}, giving $E_{C0}\simeq115~\mu$eV for both devices, while the total normal state resistances, $R_{N,L}=19~$k$\Omega$ and $R_{N,H}=18~$k$\Omega$, gave us Ambegaokar-Baratoff values \cite{anderson64} for the Josephson energies per junction of $E_{J,L}=78~\mu$eV and $E_{J,H}=82~\mu$eV.

In Figs. \ref{fig1}(b) and (c) we show $I_{sw}(n_{g})$ curves for the L and H devices measured at $T=30$~mK with a ramp rate $\dot{I}=1~\mu$A/s (1~nA/ms). The devices behave quite differently: the L device is $1e$ periodic while the H device is $2e$ periodic. This behavior has been observed in eighteen devices made in three different fabrication sessions but having very similar values of $\Delta_{i}$, $\Delta_{\ell}$, $E_{C0}$, and $R_{N}$. It was also seen in measurements of the zero-bias phase diffusion resistance (not shown), where the devices were never driven to the voltage state. Since thermal QPs are strongly frozen out, the $1e$ period in the L device implies a source of nonequilibrium QPs. Presumably, the co-fabricated H device contains a similar QP source, but its $2e$ period seems to indicate that these QPs do not enter the island. Although this might be naively expected from the barrier-like gap profile of the H device, we will show below that the actual situation is more complex. We begin by constructing a model that explains when nonequilibrium QPs in the leads should enter the island of an SCPT.

\begin{figure}[tbp]
\includegraphics[scale=0.45]{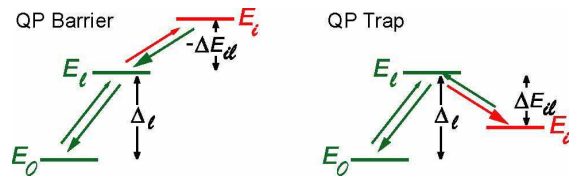} %0.5
\caption{Three-state model of poisoning. Transitions from 0 to $\ell$ are driven by a nonequilibrium QP source in the leads. When $\Delta E_{\ell i}<0$ these QPs see the island as a barrier and the parity is predominantly even. When $\Delta E_{\ell i}>0$ QPs see the island as a trap and parity is predominantly odd at low temperatures, but QPs can be thermally activated out of the island at higher temperatures.}
\label{fig2}
\end{figure}

We consider here three states: the fully paired (no QPs) ``0'' state, the $i$ state with one QP on the island, and the ``$\ell$'' with one QP in the leads, near a junction. We assume QPs cannot be spontaneously created on the island, thus poisoning is a two-step process: a QP is created in the leads (transition $0\rightarrow\ell$, parity remains even) and then tunnels onto the island (transition $\ell\rightarrow i$, parity switches to odd). From the $\ell$ state, the QP may also diffuse away from the junction or recombine with another QP (transition $\ell\rightarrow0$), in which case poisoning does not occur. For the reduced range $n_{g}:0\rightarrow1$, the energy of the 0 state is the ground state energy of the Coulomb and Josephson components of the Hamiltonian (see, for instance, Ref.~\cite{joyezthesis}), $E_{C\!-\!J}^{n=0}(n_{g},\varphi)$, where $\varphi$ is the superconducting phase difference across the device. The energies of the $\ell$ and $i$ states are $E_{\ell} = E_{C\!-\!J}^{n=0}(n_{g},\varphi) +  \Delta_{\ell}$ and $E_{i} = E_{C\!-\!J}^{n=1}(n_{g},\varphi) +  \Delta_{i}$. The energy change for $0\leftrightarrow\ell$ transitions is a constant, $\Delta_{\ell}$. The energy change for parity-switching $\ell\leftrightarrow i$ transitions is 
\begin{equation}
\Delta E_{\ell i}(n_{g},\varphi) \equiv E_{\ell}-E_{i}=\delta E_{C\!-\!J}^{0\rightarrow1}(n_{g},\varphi)+\delta\Delta.
\label{deleli}
\end{equation}
Figure~\ref{fig2} illustrates the energy levels of our model for two cases. When $\Delta E_{\ell i}<0$, QPs in the leads are prevented from reaching the island by a barrier, thus poisoning is suppressed. When $\Delta E_{\ell i}>0$, QPs will readily tunnel onto the island and can become trapped there. This trapping behavior means that even a weak QP source in the leads can result in strong poisoning of an SCPT.

\begin{figure}[tbp]
\includegraphics[scale=0.5]{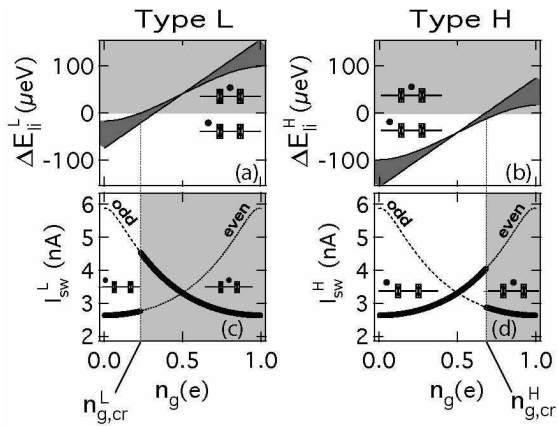} %0.5
\caption{(a)--(b) $\Delta E_{\ell i}$ for the L and H devices. The range of $\Delta E_{\ell i}$ for all values of $\varphi$ is indicated by the dark gray region. (c)--(d) Predicted crossover in $I_{sw}$ between even and odd parity. The dotted (dashed) curves are the even (odd) curves from Fig. \protect\ref{fig1}(a) while the thick lines follow the curve corresponding to the lowest energy QP state. The island traps QPs for $n_{g}>n_{g,cr}$ (light gray regions).}
\label{fig3}
\end{figure}

For $|\delta\Delta|<\delta E_{C\!-\!J}^{0\rightarrow1}$, our model predicts that a single device can span both barrier and trap regimes as $n_{g}$ is varied, regardless of gap profile. Our experiments test this prediction quantitatively. During each switching measurement, $n_{g}$ is fixed but, due to phase diffusion, $\varphi$ varies rapidly as the bias current is ramped \cite{kautzprb}. The upper part of Figure~\ref{fig3} shows $\Delta E_{\ell i}$ $\it{vs.}$ $n_{g}$ with the range of values due to variations in $\varphi$ indicated by a dark gray band. The point at which $\Delta E_{\ell i}$ first crosses zero for any value of $\varphi$ defines the critical value $n_{g,cr}$ above which the system can trap QPs. The lower part of Fig.~\ref{fig3} shows the corresponding $I_{sw}(n_{g})$ curves predicted by the model assuming the system always occupies the lowest energy QP state. For our devices, $n_{g,cr}^{L}=0.23$ and $n_{g,cr}^{H} = 0.68$ (with uncertainties of $\sim$20\%).

\begin{figure}[tbp]
\includegraphics[scale=0.4]{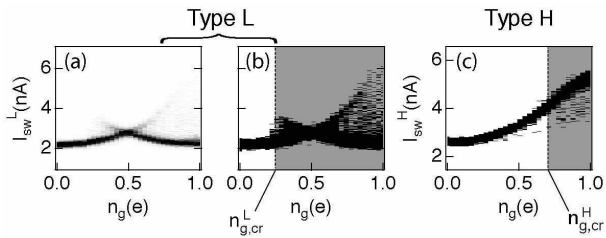} %0.8
\caption{(a) $I_{sw}(n_{g})$ for the L device ramped at $\dot{I}=100~\mu$A/s and $T=30$~mK. The grayscale is saturated at half the maximum count number to emphasize rare events. (b) Same as (a) but with all switching events shown equally. The gray region marks the predicted range for QP trapping. (c) $I_{sw}(n_{g})$ for the H device ramped at $\dot{I}=1~\mu$A/s with all counts shown equally. (We show data for the slower ramp rate to emphasize events that occur \textit{below} the dominant $2e$ curve.)}
\label{fig4}
\end{figure}

To compare our data with the predictions above we must consider parity fluctuations due to finite temperature (or external noise) that can cause the system to occupy excited states. The ability of our measurement to resolve parity fluctuations is determined by the fluctuation rates $\Gamma_{\ell i(i\ell)}$ for lead-to-island (island-to-lead) QP tunneling and the time to ramp between the two switching currents at a given $n_{g}$, $\tau _{ramp}\equiv |\langle I_{sw}^{even}\rangle -\langle I_{sw}^{odd}\rangle |/\dot{I}$. When $\tau _{ramp}\gg\Gamma_{\ell i(i\ell)}^{-1}$ only the lower critical current at a given $n_{g}$  is observed since fluctuations into this state occur before the ramp can reach the higher critical current. It is important to note that these fluctuations will occur even if the occupation probability of the state with lower critical current is small. Thus a slow enough ramp will only reveal the smaller of $I_{sw}^{even}$ and $I_{sw}^{odd}$ at a given $n_{g}$, yielding perfect $1e$ periodicity as we observe for the L device (Fig.~\ref{fig1}(b)). In Figure~\ref{fig4}(a) we shorten $\tau_{ramp}$ by a factor of 100 and observe that \emph{both} even and odd states are populated, indicating that the parity is indeed fluctuating. Since we do not observe both states until we ramp at this rate, we estimate $\Gamma_{\ell i(i\ell)}^{-1s}\sim10~\mu$s--- too fast to see with our slower ramps. When we show all counts for $I_{sw}(n_{g})$ equally in Fig.~\ref{fig4}(b), rather than using a grayscale, we see that the odd state is only occupied when $n_{g}\gtrsim0.2$, in good agreement with the value $n_{g,cr}^{L}=0.23$ derived above.

Parity fluctuations are also apparent in the H device when we plot all counts equally (Fig.~\ref{fig4}(c)). The small number of counts below $I_{sw}^{even}$ indicates occasional fluctuations to the odd state, occuring for $n_{g}\gtrsim0.7$ as predicted \cite{tau_run}. This demonstrates that the H device is in fact poisoned, despite the clean $2e$ curve in Fig.~\ref{fig1}(c), illustrating how a grayscale or average plot of $I_{sw}(n_{g})$ can fail to reveal QP poisoning. Comparing Figs.~\ref{fig4}(b) and (c), we see that the poisoning is considerably weaker in the H device than in the L device. Furthermore, the temperature dependence of the two devices is very different: $I_{sw}(n_{g})$ of the L device changes considerably as $T$ is raised (as discussed below) while the H device behavior remains unchanged until $T\gtrsim300~$mK when thermal QPs begin to populate the island. To explain these aspects of our data we return to the model shown in Fig.~\ref{fig2}.

%We assume $0\rightarrow\ell$ transitions are driven by a nonequilibrium QP source while $\ell\rightarrow0$ transitions occur either by recombination or diffusion to normal metal in the leads. We thus approximate the corresponding rates $\Gamma_{0\ell(\ell0)}$ as independent of $T$. Assuming $\ell\leftrightarrow i$ transitions are thermally activated, the corresponding rates are $\Gamma_{i\ell}=A\exp(-\Delta E_{\ell i}/kT)$ and $\Gamma_{\ell i}=A$ for $\Delta E_{\ell i}>0$ and $\Gamma_{\ell i}=A\exp(-|\Delta E_{\ell i}|/kT)$ and $\Gamma_{i\ell}=A$ for $\Delta E_{\ell i}<0$. From detailed balance, the steady state ratio between the probability of the even ($p_{even}=p_{0}+p_{\ell}$) and odd ($p_{odd}=p_{i}$) states is

\begin{figure}[tbp]
\includegraphics[scale=0.45]{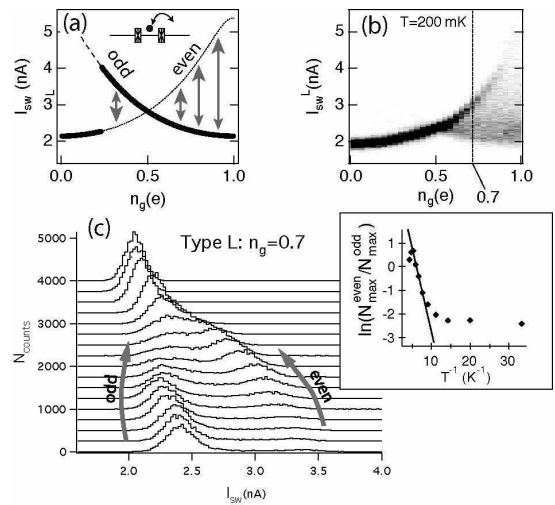}
\caption{(a) Thermal activation of QPs out of the island of the L device in the trapping regime. (b) $I_{sw}(n_g)$ for the L device at 200 mK. (c) Histograms of $I_{sw}$ at $n_{g}=0.7$ for the L device as a function of temperature. The histograms span $T=30$~mK to 350~mK in 20~mK increments (from bottom to top) and are offset vertically for clarity. \textit{Inset:} Peak height ratio and fit to thermal activation model (see text).}
\label{fig5}
\end{figure}

Since our model is based on a nonequilibrium QP source, we assume the rates $\Gamma_{0\ell(\ell0)}$ for $0\rightarrow\ell$ ($\ell\rightarrow0$) transitions are approximately independent of $T$. Assuming $\ell\leftrightarrow i$ transitions are thermally activated, the ratio of the corresponding rates is $\Gamma_{i\ell}/\Gamma_{\ell i}=\exp(-\Delta E_{\ell i}/kT)$.  From detailed balance, the steady state ratio between the probability of the even ($p_{even}=p_{0}+p_{\ell}$) and odd ($p_{odd}=p_{i}$) states is
\begin{equation}
\alpha\equiv\frac{p_{even}}{p_{odd}}=\left(1+\frac{\Gamma_{\ell0}}{\Gamma_{0\ell}}\right)\frac{\Gamma_{i\ell}}{\Gamma_{\ell i}}=\alpha_{0\ell} e^{-\Delta E_{\ell i}/kT},
\label{thermact}
\end{equation}
where $\alpha_{0\ell}\equiv(1+\Gamma_{\ell0}/\Gamma_{0\ell})$. Equation~\ref{thermact} predicts the following for an SCPT with $\Delta E_{\ell i}>0$. If there are no QPs in the leads, $\Gamma_{0\ell}=0$ and the device will always be in the even state, as expected. However, for any nonzero $\Gamma_{0\ell}$ the device will become trapped in the \emph{odd} state as $T\rightarrow0$. Furthermore, as $T$ is raised QPs can be thermally activated out of the trap, and for $T>T^*=\Delta E_{\ell i}/k\ln{(\alpha_{0\ell})}$ the device will be predominantly in the \emph{even} state (assuming thermal QPs are still negligible). The effect of this process on $I_{sw}(n_{g})$ is shown schematically in Figure~\ref{fig5}(a).

With thermal activation, our model explains both L and H devices. In the L device $I_{sw}(n_{g})$ changes little below 100~mK, suggesting that the effective temperature of QPs on the island does not change below this temperature. However, above 100~mK, the occupation probability shifts rapidly towards the even state and, at 200~mK, the even state is seen more often than the odd state (see Fig.~\ref{fig5}(b)). In Fig.~\ref{fig5}(c) we display histograms of $I_{sw}$ for the L device at $n_{g}=0.7$. The histograms show peaks corresponding to the even/odd states and their evolution with $T$. To first order we can take the ratio of the peak heights, $N_{max}^{even}/N_{max}^{odd}$, as a measure of $\alpha$. The inset of Fig.~\ref{fig5}(c) shows that this ratio increases exponentially from 100~mK to 200~mK (the peaks become difficult to distinguish above 200~mK).  A fit using Eqn.~\ref{thermact} yields a trap depth $\Delta E_{\ell i}^{fit}=62\pm3~\mu$eV that agrees with the value of $72\pm13~\mu$eV predicted by Equation~\ref{deleli}. The fit also yields $\alpha_{0\ell}^{fit}\sim100$, indicating that QPs occupy the $\ell$ state from which poisoning can occur $\sim$1\% of the time. We note that behavior consistent with this thermal activation picture was found in at least one other study of SCPTs \cite{vanderwalthesis}.

Using the prefactor $\alpha_{0\ell}^{fit}$ obtained above, we can predict the level of poisoning in the H device. For the maximum trap depth $\Delta E_{\ell i}(n_{g}=1)=17~\mu$eV, Eqn.~\ref{thermact} predicts $T^{*}=40$ mK. At the minimum effective temperature of 100~mK reached in the L device, we expect $\alpha\geq10$ at $n_{g}=1$, \textit{i.e.}, $p_{even}\sim1$. Thus the parity of the H device remains mostly even because its gap profile makes $\Delta E_{\ell i}$ small and QPs do not become cold enough to remain on the island with such a shallow trap.

In summary, we have demonstrated control over the gap profile in SCPT transistors and studied QP poisoning in devices with two types of profiles. We find that the behavior of these devices can be quite complex, depending on gap profile, gate voltage, temperature, and measurement timescales. A three-state, nonequilibrium QP model correctly predicts the range of gate voltage over which QPs can be trapped on the island and the temperature dependence of our devices. It also predicts that a weak QP source can cause strong poisoning, unless the trapping regime can be eliminated entirely by making a type H device with a larger $\delta\Delta$. Finally, our H device demonstrates that the traditional method for detecting QP poisoning in SCPTs is not completely reliable.  Applications such as metrology and quantum computing in which QP free operation is critical may require better tests of QP poisoning, perhaps involving realtime detection of individual tunneling events. 

%\begin{acknowledgments}
The authors would like to acknowledge useful discussions with D. Esteve.
%\end{acknowledgments}

%\bibliography{qp2}
%ESUBS WILL ONLY TAKE ONE TEXT FILE!
%CUT AND PASTE *.bbl file here:

\end{document}